\documentclass[isre,nonblindrev]{informs3}

\DoubleSpacedXI 

\usepackage{natbib}
 \bibpunct[, ]{(}{)}{,}{a}{}{,}%
 \def\bibfont{\small}%
\usepackage{bbm}

\TheoremsNumberedThrough     

\EquationsNumberedThrough    

\MANUSCRIPTNO{ISR-2019-494} 

\begin{document}


\RUNAUTHOR{Holtz and Aral}

\RUNTITLE{Limiting Bias from Test-Control Interference}

 \TITLE{Limiting Bias from Test-Control Interference in Online Marketplace Experiments}

\ARTICLEAUTHORS{%
\AUTHOR{David Holtz, Sinan Aral}
\AFF{MIT Sloan School of Management, \EMAIL{dholtz@mit.edu}, \EMAIL{sinan@mit.edu}}
} 

\ABSTRACT{%
Many internet firms use A/B tests to make product decisions. In an A/B test, the typical objective is to measure the total average treatment effect (TATE), which measures the difference between the average outcome if all users were treated and the average outcome if all users were untreated. However, a simple difference-in-means estimator will give a biased estimate of the TATE when outcomes of control units depend on the outcomes of treatment units, an issue we refer to as test-control interference. Practically speaking, this type of interference is quite common in online marketplaces and failure to account for it can lead to estimates that are inflated by a factor of 100\%. Using a simulation built on top of data from Airbnb, this paper considers the use of methods from the network interference literature for online marketplace experimentation. We model the marketplace as a network in which an edge exists between two sellers if their goods substitute for one another. We then simulate seller outcomes, specifically considering a ``status quo" context and ``treatment" context that forces all sellers to lower their prices. We use the same simulation framework to approximate TATE distributions produced by using blocked graph cluster randomization, exposure modeling, and the Hajek estimator for the difference in means. We find that while blocked graph cluster randomization reduces the bias of the naive difference-in-means estimator by as much as 62\%, it also significantly increases the variance of the estimator. On the other hand, the use of more sophisticated estimators produces mixed results. While some provide (small) additional reductions in bias and small reductions in variance, others lead to increased bias and variance. Overall, our results suggest that experiment design and analysis techniques from the network experimentation literature are promising tools for reducing bias due to test-control interference in marketplace experiments.
}%



\maketitle

\section{Introduction}

A common way for internet firms to make product decisions is through experimentation, or `A/B testing.' Typically, users who visit a website during an A/B test are randomly assigned to either the test or control group. These two groups experience different variants of the online product. After some amount of time, a `winner' is chosen by evaluating the variants' relative performance. This evaluation is performed by calculating the total average treatment effect (TATE) for a set of predetermined metrics (e.g., click-through rate, average revenue per user) and performing a statistical significance test \citep{kohavi2009controlled}. Conceptually, the TATE is meant to measure the difference in average outcome across all units between the counterfactual situation in which 100\% of units are exposed to the treatment, and the status quo situation in which 0\% of units are exposed to the treatment.

Under the assumption that each experimental unit's response is not influenced by any other units' treatment assignment, the TATE can be identified by randomly assigning units to treatment or control, and calculating the difference in means between the two groups. This assumption is often referred to as the `stable unit treatment value assumption' (SUTVA) \citep{rubin1974estimating} or the `no interference' assumption \citep{cox1958planning}. However, online marketplaces are by definition connected. For instance, a treatment that increases buyer demand may reduce the supply available to buyers in the control. Similarly, a treatment that induces sellers to lower prices may increase demand for treatment sellers' products, thereby reducing the demand for control sellers' products. In both of these cases, the treatment affects not only the treated units' outcomes, but also the control units' outcomes.

In the online experimentation literature, this phenomenon is referred to as `test-control interference.'  \citet{blake2014marketplace} find empirical evidence for test-control interference in marketplaces by analyzing an email marketing experiment performed on eBay. They conclude that naive estimates of the TATE ignoring test-control interference exaggerate the treatment's effectiveness by about a factor of 2. Using a simulation based approach, \citet{fradkin2014search} finds that search algorithm experiments in online marketplaces can overstate true treatment effects by over 100\%. 

Some of the world's most valuable tech companies, such as Alibaba, Amazon, Airbnb, and Uber, own and operate online marketplaces. These firms conduct thousands of experiments a year \citep{clarke2016}, and often hope to measure treatment effects that are small in relative terms (e.g., fractions of a percentage point), but which can correspond to large revenue gains or losses (e.g., millions of dollars). Given the high potential cost of drawing incorrect inferences about the effects of any given product change, it is crucial for these firms to develop methods for unbiased causal inference in online marketplace settings.

Both \citet{blake2014marketplace} and \citet{fradkin2014search} propose methods for combatting test-control interference in marketplace experiments. Rather than compare outcomes across buyers, \citet{blake2014marketplace} limit their analysis to auctions where a majority of bidders are in the treatment or control, and compare outcomes across auctions. While this does eliminate within-auction interference between users, there is still potential bias in their effect size estimates due to interference across auctions. Furthermore, it is not clear how well this methodology generalizes to marketplaces that do not offer convenient units of analysis (e.g., auctions) over which to aggregate outcomes. \cite{fradkin2014search} proposes randomizing across well-defined markets (rather than buyers or sellers), or combining experimental results with results from complicated structural model simulations. However, market level experiments are often infeasible (due to ambiguous - or non-existent - market definitions, a limited number of markets, or market heterogeneity), and the simulations necessary to quantify equilibrium treatment effects may be too complicated and/or cumbersome for many firms or researchers to implement.

In this paper, we use simulated online marketplace experiments to quantify the extent to which methods from the network experimentation literature, such as graph cluster randomization \citep{ugander2013graph}, exposure modeling \citep{aronow2012estimating}, and treatment effect estimators such as the Hajek estimator \citep{hajek1971comment} can reduce bias arising from test-control interference in online marketplaces when applied to product networks. While previous research has shown that these methods are effective in more traditional ``networked" settings, it is not obvious \textit{ex ante} that they will be effective at reducing bias in treatment effect estimates in online marketplace settings, given that the underlying phenomena that cause spillovers (e.g., substitution and complementarity) are different, and that there is often not an explicit and/or definitive ``network" that connects items or sellers in a marketplace.

Prior research in the information systems literature has shown that ``product networks" can be an effective tool to study competition and demand spillovers in online marketplaces. \citet{Oestreicher-Singer2012} empirically test the hypothesis that a visible co-purchase or co-view link between two books on Amazon.com increases their demand correlation, and find that such links lead to a threefold increase in the influence that complementary products have on each others' demand levels. In a later paper, \cite{Oestreicher-Singer2012a} find that the extent to which peer-based recommendations in online marketplaces redirect demand to ``niche" products in the long tail depends on the network structure of co-purchase or co-view ``edges" between items in a given category. Extending this research, \citet{Dhar2014} find that when time series data is available, product networks can be used for demand prediction. While each of these papers focuses on ``visible" product networks, in which an edge exists between two products if they appear on each other's checkout pages, in this paper we build a product network in which edges are based on attributes that items share, such as price point, category, or physical location.


We begin with a scraped dataset consisting of Airbnb properties and their attributes. Using a simple decision rule that defines which listings are likely to substitute for one another, we infer an underlying network structure: an edge exists between two listings if they are likely to substitute for one another. We next simulate a sequential process of searchers with heterogeneous preferences visiting Airbnb, facing a consideration set provided by a search algorithm, and then making a purchase decision.  Using our simulation framework, we measure the booking rate and the average revenue per listing both in the baseline case and under policy changes that require all listings to lower their price. 

Having quantified the true TATEs via simulation, we conduct simulated experiments and compare the bias and RMSE of the samplings distributions corresponding to the difference-in-means estimator with independent randomization, the difference-in-means estimator with graph cluster randomization and cluster-level block random assignment, graph cluster randomization and block-level random assignment along with the fractional neighborhood treatment response (FNTR) exposure model, and graph cluster randomization and block-level random assignment along with the FNTR exposure model and the Hajek estimator. Notably, our approach does not make any assumptions about the functional form of interference between units. In fact, our marketplace simulations do not explicitly prescribe that marketplace sellers interfere with one another at all.

We find that blocked graph cluster randomization reduces the bias of the difference-in-means estimator by as much as 62\%. However, it also increases the RMSE of the difference-in-means estimator by as much as 75\%. Although analysis techniques such as exposure models and the Hajek estimator do in some cases further reduce bias, the bias reductions we observe are much smaller than those attributable to blocked graph cluster randomization. Furthermore, these analysis techniques do not generally reduce the RMSE of the treatment effect estimator. Our results show that the methods we propose in this paper can reduce the bias of experimental treatment effect estimates in online marketplace settings, but these bias reductions come at the cost of significant increases to the variance of the treatment effect estimator's sampling distribution. This finding highlights a critical tradeoff for marketplace designers and other practitioners: while unbiased experiments are desirable, the corresponding loss of statistical power can lead to minimum detectable effect sizes that are much higher than is acceptable. This tension is particularly important given that treatment effects that are small in relative terms can equate to gains or losses of millions of dollars for many online marketplace firms.

The structure of this paper is as follows. In Section \ref{theory}, we review the experiment designs and analysis methods to be evaluated via simulation. Section \ref{data_desc} provides a summary of the Airbnb dataset used, the network generation process, and our approach for clustering the network. We describe the simulation process in Section \ref{simulations}, and compare the bias and variance achieved with different experiment designs and analysis techniques in Section \ref{results}. Section \ref{theory_extensions} discusses our findings and describes ways to further this work. In Section \ref{conclusions}, we conclude.

\section{Theoretical Motivation} \label{theory}

We first motivate and describe the experiment design and analysis methods considered in this paper. Suppose we have a network with $N$ units. Define $Z$ as a vector of length $N$ indicating each unit's treatment assignment. Now, define $Y_i(Z=z)$ as unit $i$'s outcome when the global treatment vector $Z$ is equal to $z$. The total average treatment effect is defined as 

\begin{equation}
\tau(z_1, z_0) = \frac{1}{N} \sum_{i=1}^N \mathbb{E} \left [ Y_i(z_1) - Y_i(z_0) \right ],
\end{equation}

\noindent where $z_1$ denotes a treatment assignment vector in which every unit is assigned treatment, and $z_0$ denotes a treatment assignment vector in which every unit is assigned control. Since, during any given experiment, it is impossible to simultaneously observe the counterfactual in which all units are treated and the counterfactual in which all units are untreated, researchers often assume independence between units (SUTVA). Under this assumption, $Y_i(Z)$ only depends on $Z_i$. As a result, $Y_i(Z)$ can be simplified to $Y_i(Z_i)$, and the TATE is equivalent to the average treatment effect (ATE). In conjunction with random assignment, this assumption typically allows $\tau$ to be identified by a simple difference-in-means estimator. However, when SUTVA is violated, as in the case of online marketplaces, the difference-in-means estimator will be biased.

One method to reduce this bias is to randomize units in such a way that they are ``closer" to the global treatments of interest. \citet{ugander2013graph} and \citet{eckles2014design} show this can be achieved via graph cluster randomization. In the case of independent assignment, each $Z_i$ is determined by drawing from a Bernoulli distribution,

\begin{equation}
Z_i \sim \textrm{Bernoulli}(q),
\end{equation}

\noindent where $q$ is the probability of assignment to the treatment. In the case of graph cluster randomization, the network is partitioned into $N_C$ clusters: $C_1, C_2, ...., C_{N_C}$. After the network is partitioned, each cluster is assigned a treatment, $W_j$, by drawing from a Bernoulli distribution,

\begin{equation}
W_j \sim \textrm{Bernoulli}(q),
\end{equation}

\noindent where again, $q$ is the probability of cluster assignment to treatment. Each vertex in a given cluster is then assigned its cluster's treatment assignment. 

Another way to reduce bias is to propose an ``exposure model" that dictates the way that a unit's neighbors' treatment assignments impacts that unit's outcomes. Once an exposure model is specified, it is possible to define a function $g_i(\cdot)$ such that $g_i(z_m) = g_i(z_n)$ implies that treatment vectors $z_m$ and $z_n$ provide unit $i$ with the same ``effective treatment" \citep{manski2013identification}. Once an effective treatment (and effective control) are defined, the TATE estimator can be modified so that only units exposed to the effective treatment or effective control are included in the analysis:

\begin{equation}
\tau(z_1, z_0)_{eff} = \frac{1}{N} \sum_{i=1}^N \mathbb{E} \left [ Y_i(Z) | g(Z) = g(Z_1) \right ] - \mathbb{E} \left [ Y_i(Z) | g(Z) = g(Z_0) \right ].
\end{equation}

\noindent In this paper, we consider the fractional neighborhood treatment response (FNTR) exposure model discussed in \citet{eckles2014design} and \citet{ugander2013graph}. In the FNTR exposure model, a vertex is in the effective treatment (control) if it is treated (not treated) and more than some fraction $\lambda$ of its neighbors are in the treatment (control).\footnote{We consider $\lambda = .50$, $\lambda = .75$ and $\lambda = .95$.}

Under an exposure model, the difference in means estimator will only be unbiased if, for the randomization process in question,

\begin{equation}
E \left [ Y_i | g_i(Z) = g_i(z) \right ] \perp \mathbb{P}\left [ g_i(Z) = g_i(z) \right ].
\end{equation}

\noindent In other words, the expected outcome for unit $i$ under a particular effective treatment is independent of the probability that unit $i$ receives that effective treatment. However, this assumption will often fail when units are assigned to different treatment arms using graph cluster randomization, and/or when exposure models are used for analysis. For instance, one could imagine that higher degree nodes (in this context, items that are similar to many others) might have a different expected outcome than lower degree nodes (in this context, very unique items), even conditional on being exposed to the same treatment. It may also be the case that a higher degree node is less likely to be assigned to the effective treatment or control, since that unit has a) more edges and b) potentially more edges that span multiple clusters. The Horvitz-Thompson estimator \citep{horvitz1952generalization},

\begin{equation}
\hat{\tau_{HT}}(z_1, z_0) = \frac{2}{N}  \left ( \sum_{i=1}^N \frac{Y_i \mathbbm{1} \left [ g_i(Z) = g_i(z_1) \right ]}{\pi_i(z_1)} -  \sum_{i=1}^N \frac{Y_i \mathbbm{1} \left [ g_i(Z) = g_i(z_0) \right ]}{\pi_i(z_0)} \right )
\end{equation}

\noindent can provide unbiased estimates by reweighting observations according to their probability of being assigned to different treatment conditions. However, this estimator often exhibits high variance. As a solution to this issue, both \citet{aronow2012estimating} and \citet{eckles2014design} use the Hajek estimator \citep{hajek1971comment} as defined by:

\begin{equation}
\begin{split}
\hat{\tau_H}(z_1, z_0) = \left ( \sum_{i=1}^N \frac{\mathbbm{1} \left [ g_i(Z) = g_i(z_1) \right ]}{\pi_i(z_1)} \right )^{-1} \sum_{i=1}^N \frac{Y_i \mathbbm{1} \left [ g_i(Z) = g_i(z_1) \right ]}{\pi_i(z_1)} - \\ \left ( \sum_{i=1}^N \frac{\mathbbm{1} \left [ g_i(Z) = g_i(z_0) \right ]}{\pi_i(z_0)} \right )^{-1} \sum_{i=1}^N \frac{Y_i \mathbbm{1} \left [ g_i(Z) = g_i(z_0) \right ]}{\pi_i(z_0)},
\end{split}
\end{equation}

\noindent where $\pi_i(z) = \mathbb{P}\left [ g_i(Z) = g_i(z) \right ]$ is the probability that a given unit is assigned to the effective treatment ($z_1$) or effective control ($z_0$).  \citet{aronow2012estimating} point out that the Hajek estimator can often lead to efficiency gains, but this comes at the cost of finite sample bias and more complicated variance estimation.

Many of the randomization schemes and treatment effect estimators heretofore discussed can increase the variance of TATE estimates. As \citet{gerber2012field} point out, cluster random assignment (e.g., graph cluster randomization) can lead to large increases in ATE standard errors if the cluster-level mean outcome values exhibit high levels of variability. One way to offset the loss of precision due to graph cluster randomization is to perform block random assignment \citep{gerber2012field, moore2012multivariate}. In this paper, we consider the matched pair design. Under the matched pair design, each block consists of only 2 units. Because under graph cluster randomization, treatment is assigned at the cluster level, we also block at the cluster level. Within each block, exactly one cluster is assigned to the treatment and exactly one cluster is assigned to the control. 

\section{Data \& Network Construction} \label{data_desc}

\begin{figure}[ht]
\begin{center}
\includegraphics[scale=.25]{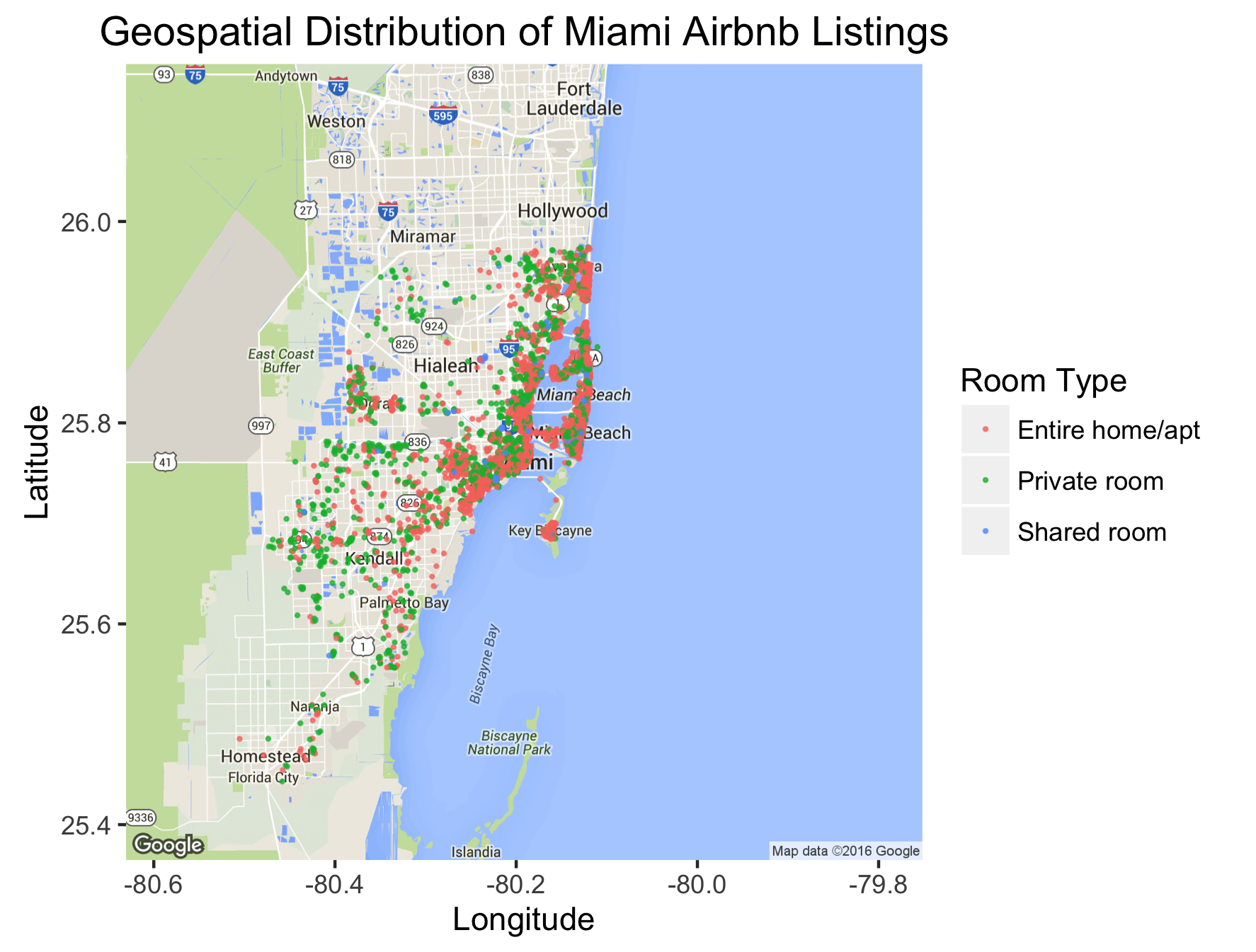}
\caption{The geospatial distribution of Airbnb listings in and around Miami. Color corresponds to listing type. This figure was produced with ggmap \citep{ggmap}.}
\label{fig:listings_map}
\end{center}
\end{figure}

We build our simulations on top of data scraped from \url{www.Airbnb.com} by \citet{slee_2015}. This dataset details the room type, number of reviews, average `overall satisfaction' rating, guest capacity, number of bedrooms, number of bathrooms, price per night (USD), minimum length of stay, latitude, and longitude of 8,855 Airbnb listings located in and around Miami. Miami serves as an ideal context for our simulation studies because it is a large enough market to produce a relatively dense substitution network, and is representative of other major metropolitan areas on Airbnb. Additionally, the dataset is small enough that our simulations are computationally tractable. Figure \ref{fig:listings_map} depicts the geospatial distribution of the listings by room type, while Tables \ref{tab:listing_room_types} and \ref{tab:listing_cov} provide information about the distribution of listing-level covariates across our sampling of Airbnb listings.

Before running our simulations, we pre-process the dataset: missing guest capacity, bedroom, and bathroom values are imputed using the modal value for each variable. Minimum length of stay values are capped at 30, and missing minimum length of stay values are imputed using the modal value for minimum length of stay. Missing overall satisfaction values are imputed using the mean value of non-empty entries.

Each row in the resulting dataset constitutes a vertex in a network. The network is meant to approximate which listings are likely to substitute for one another when searchers are making purchasing decisions. We next generate edges between vertices. Given that edges between listings are meant to suggest that those listenings substitute for one another, we stipulate that an edge is generated between two listings when the following three criteria are be satisfied:\footnote{One could imagine using a subset of these criteria (e.g., all listings within 1 mile of each other are substitutes), or a totally unrelated criteria (e.g., listings must have co-occurred in search more than $x$ times).}

\begin{enumerate}
\item The listings are within 1 mile of each other
\item The listings have the same room type
\item The difference between the guest capacity of the two listings is not greater than 1 in absolute magnitude
\end{enumerate}

\begin{figure}[ht]
\begin{center}
\includegraphics[scale=.33]{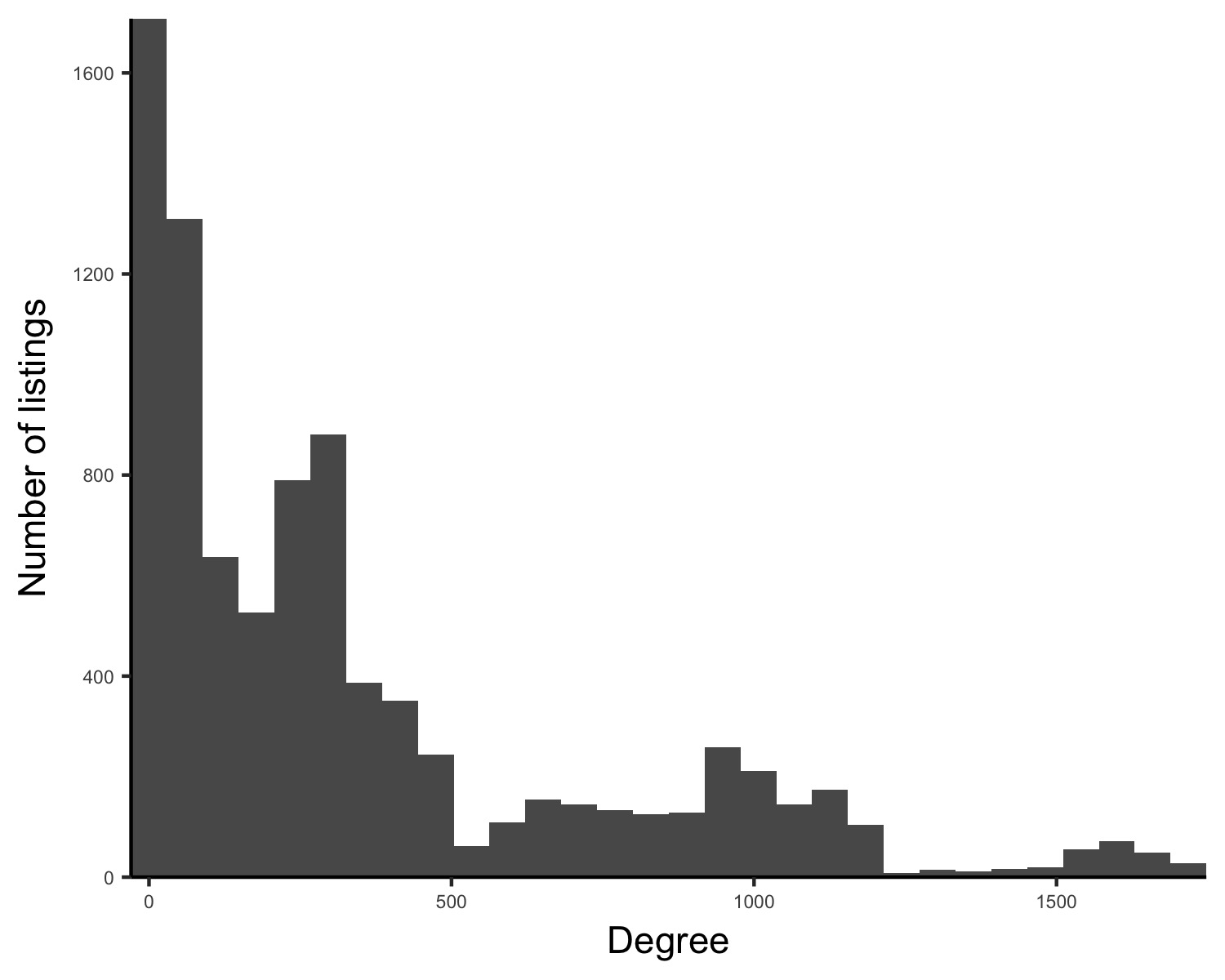}
\caption{The degree distribution for the Airbnb listing network generated using the procedure described in Section \ref{data_desc}.}
\label{fig:degree_distribution}
\end{center}
\end{figure}

\begin{figure}[ht]
\begin{center}
\includegraphics[scale=.33]{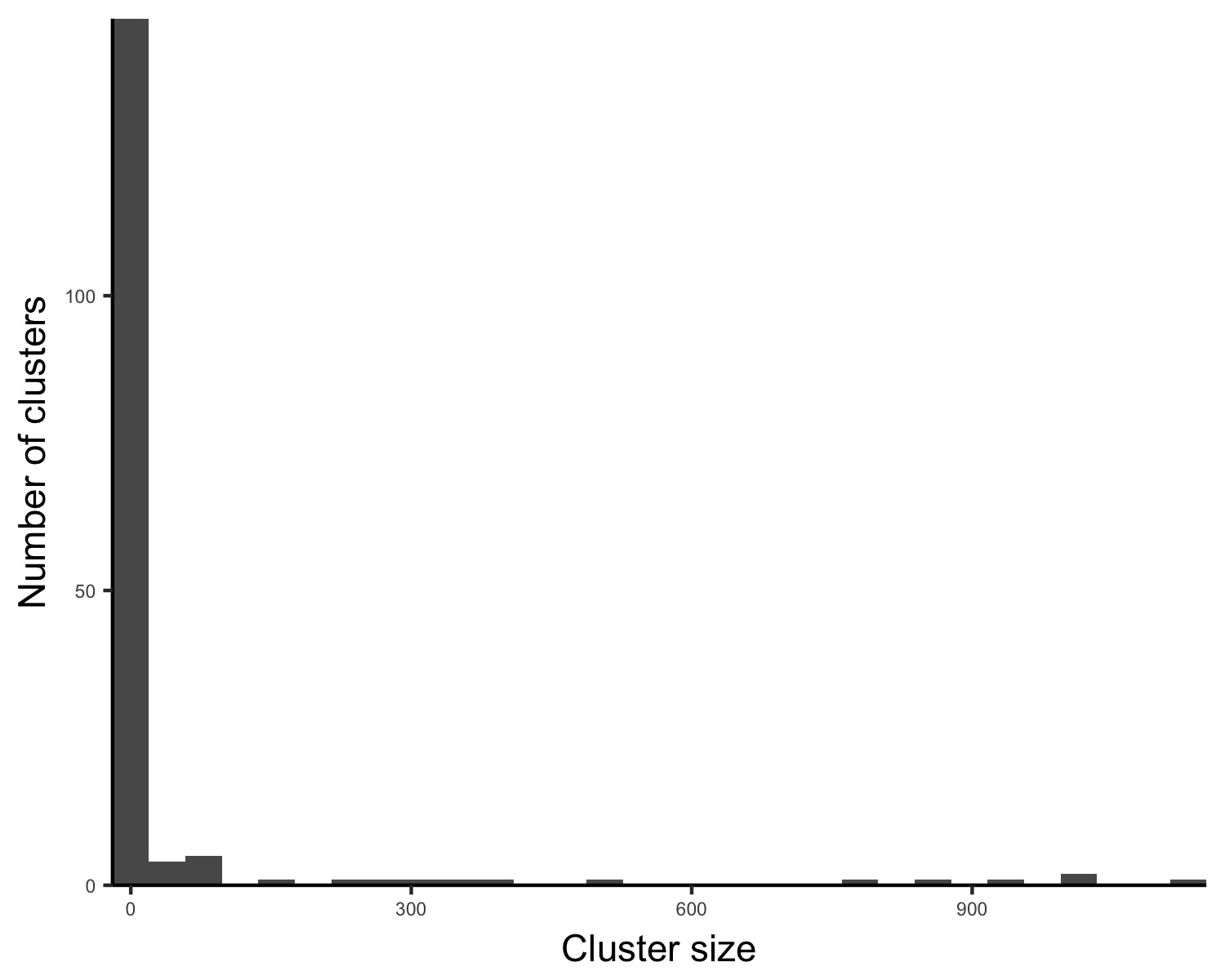}
\caption{The distribution of cluster sizes across the 169 clusters generated using Louvain clustering \citep{blondel2008fast}.}
\label{fig:cluster_size}
\end{center}
\end{figure}

The three criteria above impose a set of undirected ``substitution edges" on the set of vertices. The resulting network has 1,538,637 edges and a clustering coefficient of 0.74. Vertices in the network have an average degree of 173.76. Figure \ref{fig:degree_distribution} shows the degree distribution for the network. We proceed to divide the network into clusters using the Louvain clustering algorithm \citep{blondel2008fast}. Louvain clustering attempts to maximize modularity, a value that falls between -1 and 1 and is defined as 

\begin{equation}
Q = \frac{1}{2E} \sum_{ij} \left ( A_{ij} - \frac{d_i d_j}{2E} \right ) \mathbbm{1} ( C_i = C_j ),
\end{equation}

\noindent where $E$ is the total number of edges in the graph, $A_{ij}$ is a $\{0, 1\}$ variable that indicates whether or not an edge exists between nodes $i$ and $j$, $d_i$ and $d_j$ are the degrees of nodes $i$ and $j$, respectively, and $\mathbbm{1} ( C_i = C_j )$ is an indicator function that is equal to 1 only when $i$ and $j$ belong to the same cluster. Conceptually, Louvain clustering attempts to maximize the density of links inside communities relative to links between communities. Louvain clustering on our network leads to the network being partitioned into 169 clusters, which have an average size of 52.40 listings. Figure \ref{fig:cluster_size} shows the cluster size distribution across the 169 clusters. 

The resultant clusters are then divided into matched pairs to accommodate block random assignment. Within each cluster, the average number of reviews, average overall satisfaction score, average number of beds, average number of bathrooms, average minimum stay, average latitude, average longitude, percentage of private room listings, and percentage of shared room listings are calculated. We then calculate the Mahalanobis distance \citep{mahalanobis1936generalized} between each pair of clusters, and choose pairs to minimize the sum of the distances between clusters in the same pairs.

\section{Simulation Process} \label{simulations}

In order to estimate the true total average treatment effect of different interventions, as well as the bias and sampling variance of the TATE estimator under different experiment designs and analysis approaches, we create a framework for simulating the Airbnb booking process for one calendar night. Each set of simulated outcomes is generated using the following steps:

\begin{enumerate}
\item A `search algorithm' is randomly drawn\footnote{Randomizing over different `search algorithms' ensures that our results generalize across different ways that the firm may choose to rank their inventory, and are not due to a particular set of listing attribute weights that are strongly (mis)aligned with the approach we've used to define network edges and clusters.}
\begin{itemize}
\item The search algorithm takes into account number of reviews, overall satisfaction score, number of beds, number of baths, minimum stay length, and price
\item The search algorithm randomly draws a vector of search algorithm weights for these six attributes. The weights always sum to 1
\item The search algorithm assigns a score to each listing, computed as the weighted average (using the search algorithm weights) of centered and scaled versions of the six aforementioned variables 
\end{itemize}
\item 1,000 `searchers' sequentially look for an Airbnb listing in and around Miami. For each searcher:
\begin{itemize}
\item The searcher randomly draws a region of interest in latitude/longitude space. The locations of the box edges are drawn with uniform probability from the range of listing latitudes and longitudes. The search algorithm will only return listings in this box
\item The searcher draws a room type preference from a uniform distribution over \{entire home/apt, private room, shared room\}. The search algorithm will only return listings of this type
\item The searcher draws a minimum guest capacity from a uniform distribution over \{1,2,3,4\}. The search algorithm will only return listings with a guest capacity greater than the searcher's minimum
\item The searcher randomly draws a vector of searcher weights for the same six attributes considered by the search algorithm. Again, these weights always sum to 1
\item The searcher randomly draws a reserve score from the uniform distribution over \{0, 2\}. 
\item A searcher score is assigned to each listing. The searcher score is the weighted average (using the searcher weights) of centered and scaled versions of the six aforementioned variables
\item The search algorithm provides the searcher with a consideration set. The consideration set consists of up to 10 listings satisfying the searcher's location, room type, and guest capacity preferences with the highest search algorithm scores
\item The searcher chooses the listing in the consideration set with the highest searcher score. As long as the searcher score of that listing is above the searcher's reserve score, that listing is `booked' and can no longer appear in future searchers' consideration sets
\item If the search algorithm returns an empty consideration set or the searcher score of the highest ranked listing in the consideration set is below the searcher's reserve score, the searcher does not book any listings and exits the platform
\end{itemize}
\end{enumerate}

While our simulation framework certainly simplifies Airbnb's marketplace dynamics, we believe our work provides an effective baseline measurement of the degree to which test-control interference may bias TATE estimates in online marketplace experiments, and can help determine the extent to which the proposed experiment designs and analysis techniques can reduce that bias. 

In each of our simulations, we consider two different outcomes for Airbnb listings. The first is whether or not a listing gets booked, referred to as $Y_i$. We refer to the average of $Y_i$ over the sample as the booking rate, $BR$:

\begin{equation}
BR = \frac{1}{N} \sum_{i=1}^N \mathbb{E} \left [ {Y_i} \right ].
\end{equation}

The other outcome we consider is a listing's revenue, which we define as $R_i = Y_i p_i$. We refer to the average of $R_i$ over the sample as the average revenue per listing, $ARPL$:

\begin{equation}
ARPL = \frac{1}{N} \sum_{i=1}^N p_i \times \mathbb{E} \left [ Y_i \right ].
\end{equation}

\section{Results} \label{results}

We now use the simulation framework described in Section \ref{simulations} to measure bias in estimates of the TATE of requiring all listings to lower their price. Two different versions of the treatment are considered: one that induces listings to lower their price by 20\% (henceforth referred to as as treatment A) and one that induces listings to lower their price by 50\% (henceforth referred to as treatment B). The first treatment is more modest, whereas the second treatment should yield more pronounced test-control interference. We first use our simulation framework to estimate the ``true" TATE, i.e., the difference between the average values of our outcome variables when 100\% of listings lower their price by 20\% or 50\%, and the average values of our outcome variables when 100\% of listings experience the control. We then measure the bias in estimates of those TATEs obtained using Bernoulli randomization and the simple difference-in-means estimator. Finally, we use our simulation framework to measure the extent to which various experiment designs and analysis techniques reduce bias, and to measure the accompanying effect that they have on the RMSE of the TATE sampling distribution.

\subsection{Simulating Ground Truth}

\begin{figure}[ht]
\begin{center}
\includegraphics[scale=.33]{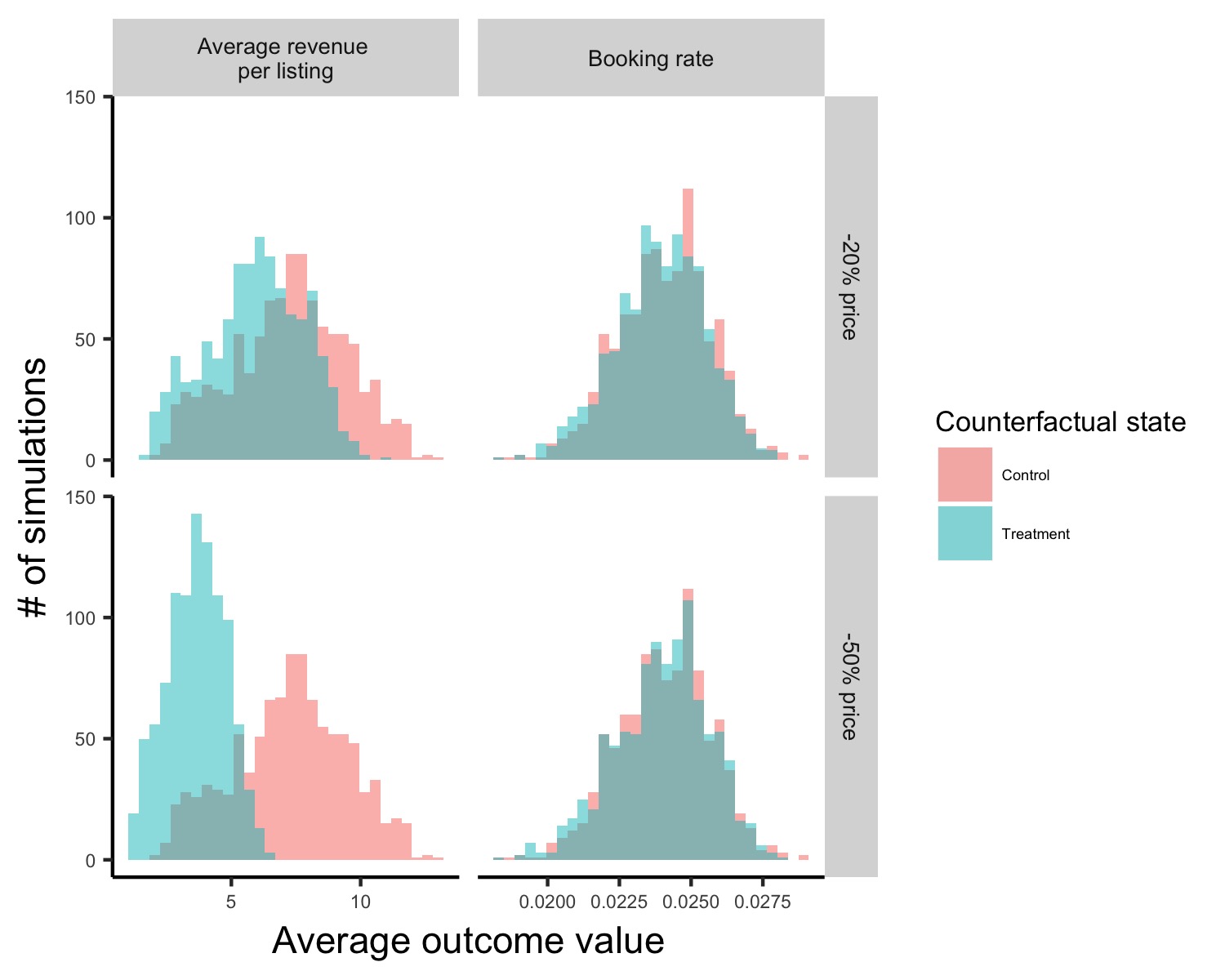}
\caption{Comparison of the simulated counterfactual average outcome distributions when either 0\% or 100\% of listings are assigned the treatment. The top row shows distributions when the treatment is a 20\% price decrease. The bottom row shows distributions when the treatment is a 50\% price decrease. The left column shows distributions for average revenue per listing. The right column shows distributions for booking rate.}
\label{fig:counterfactual_compare}
\end{center}
\end{figure}

Following the framework described in Section \ref{simulations}, we conduct 1,000 simulations of one night of booking activity for three counterfactual cases: baseline (no treatment), 100\% of listings exposed to treatment A, and 100\% of listings exposed to treatment B. For each simulation, we calculate the booking rate and average revenue per listing. Figure \ref{fig:counterfactual_compare} compares the sampling distribution of booking rate and average revenue per listing for each counterfactual case.

A two-sided $t$-test between the baseline distribution of booking rates and the counterfactual distribution of booking rates under treatment A yields a $t$-statistic of $t=2.17$ ($p = 0.03$) Although this test gives a statistically significant effect, the effect size is extremely small (TATE = $1.05 \times 10^{-5})$. This difference may arise because buyers in the baseline case exit the marketplace more often due to a lack of sufficiently ``high quality" listings. A two-sided $t$-test between the baseline distribution of booking rates and the counterfactual distribution of booking rates under treatment B yields a $t$-statistic of $t=1.12$ ($p=0.20$). In other words, we are unable to reject the null hypothesis that the booking rate in the control and the booking rate under treatment B are identical at any conventionally accepted level of statistical significance. For treatment A, while we do reject the null, the measured effect size is extremely small. In combination, these results suggest that the true effect of both treatments on the booking rate close or equal to zero.\footnote{If anything, we would expect the magnitude of the effect (if there were one) to be larger for treatment B than for treatment A.} This result is somewhat intuitive; in our simulations, regardless of the heterogeneous searcher preferences and the search algorithm that is drawn, uniformly lowering price should result in approximately the same number of bookings, whether all listings are subjected to treatment A, treatment B, or no treatment at all. Any changes in booking rate are purely a result of more or fewer searchers exiting the marketplace due to a failure to find listings above their reserve quality.

In contrast, our simulated $ARPL$ results are strikingly different. Here, two-sided $t$-tests indicate highly statistically and economically significant differences in revenue between the baseline case and either of the treatment interventions we consider. In the case of treatment A, a $t$-test between the baseline distribution of average revenue per listing and the counterfactual treated distribution of average revenue per listing yields a TATE of $-\$1.33$ ($t=14.54; p < 2.2 \times 10^{-16})$. In the case of treatment B, a $t$-test between the baseline distribution of average revenue per listing and the counterfactual treated distribution of average revenue per listing yields a TATE of $-\$3.60$ ($t=45.86; p < 2.2 \times 10^{-16}$). This result is also somewhat intuitive; assuming that the booking rate does in fact remain constant between the two treatments, the listings that do get booked after all listings have lowered their prices will make strictly less money. To get point estimates of the ``true" TATE, we take the difference in means of the counterfactual distributions. A summary of these findings can be found in Table \ref{tab:compare_baseline}.

\subsection{Estimating bias}

Having measured the ``true values" of the estimands we are interested in, we now turn our focus to measuring the bias and RMSE of the treatment effect estimator under different experiment designs and analysis techniques. In this paper, we will compare independent randomization and the difference-in-means estimator, blocked graph cluster randomization and the difference-in-means estimator, blocked graph cluster randomization with the FNTR exposure model, and blocked graph cluster randomization with the FNTR exposure model and the Hajek estimator. 

\begin{figure}[ht]
\begin{center}
\includegraphics[scale=.33]{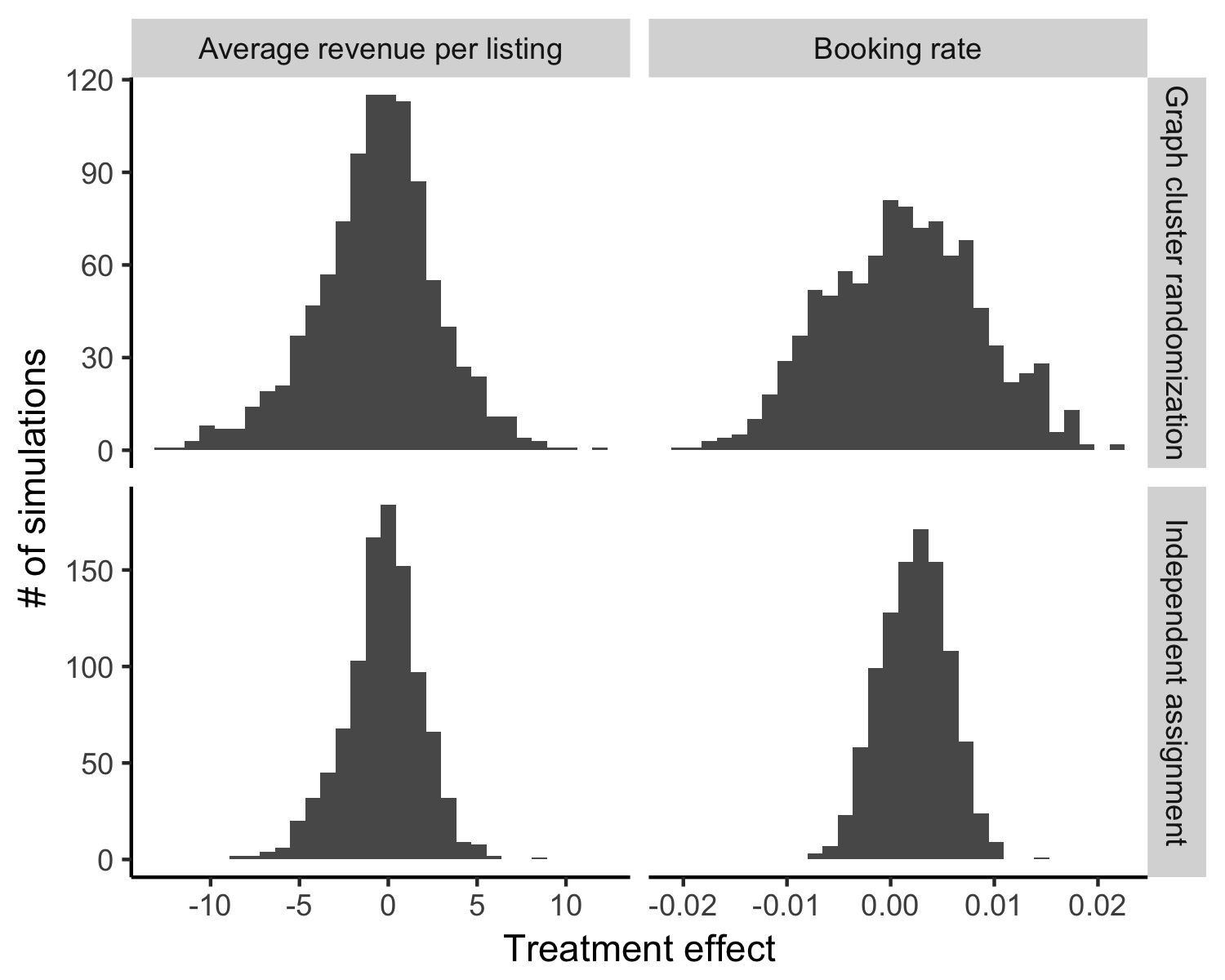}
\caption{The distribution of total average treatment effect estimates for booking rate and average revenue per listing across 1,000 simulations using a simple difference in means estimator. The treatment intervention is a 20\% reduction in listing price. The left column is the effect on average revenue per listing, whereas the right column is the effect on booking rate. The top row is the distribution when treatment is assigned via graph cluster randomization and block random assignment, the bottom row is the distribution when treatment is assigned with independent Bernoulli randomization.}
\label{fig:histogram_df_20}
\end{center}
\end{figure}

\begin{figure}[ht]
\begin{center}
\includegraphics[scale=.33]{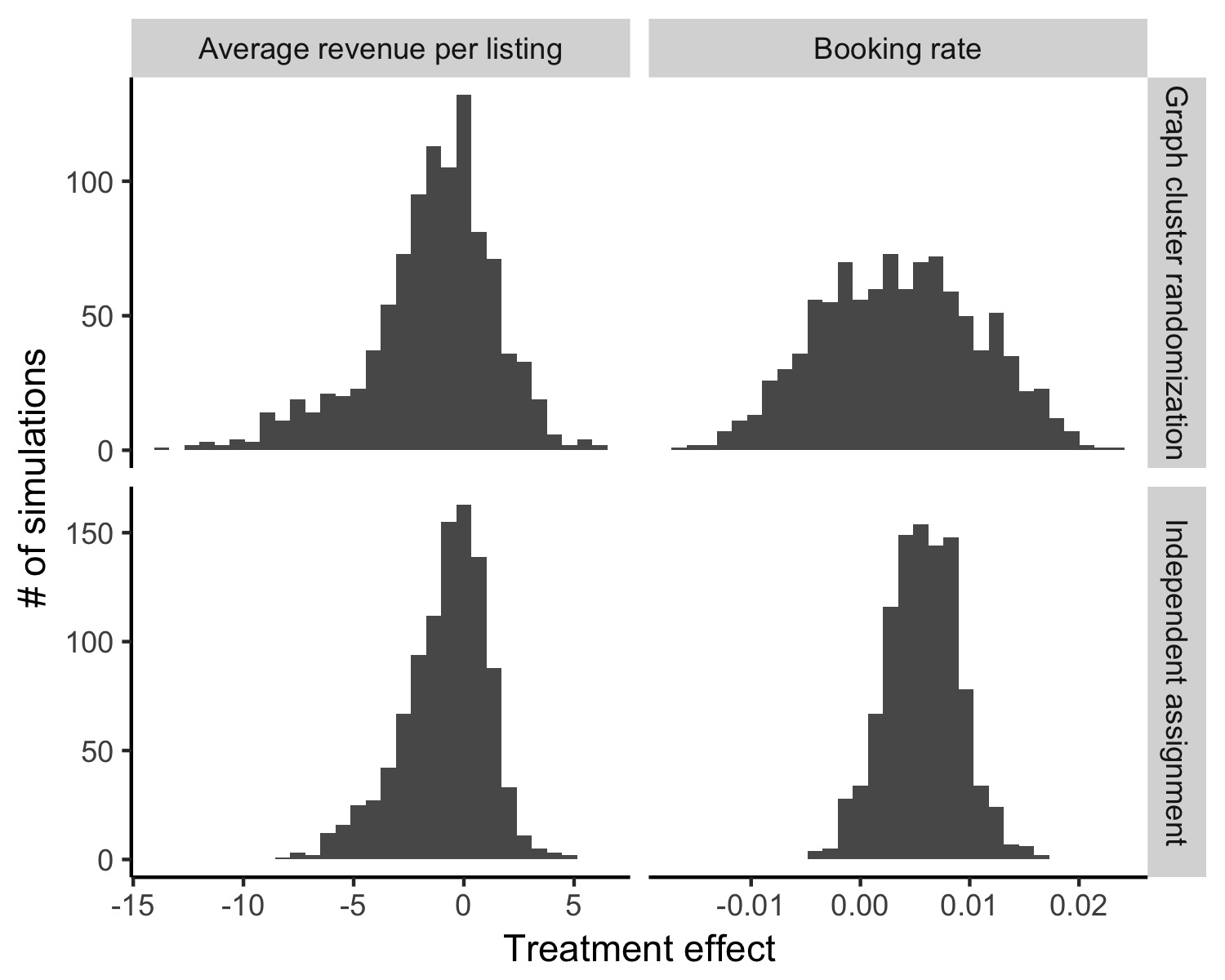}
\caption{The distribution of total average treatment effect estimates for booking rate and average revenue per listing across 1,000 simulations using a simple difference in means estimator. The treatment intervention is a 50\% reduction in listing price. The left column is the effect on average revenue per listing, whereas the right column is the effect on booking rate. The top row is the distribution when treatment is assigned via graph cluster randomization and block random assignment, the bottom row is the distribution when treatment is assigned with independent Bernoulli randomization.}
\label{fig:histogram_df_50}
\end{center}
\end{figure}

We first measure the bias and RMSE of the difference-in-means estimator when used in conjunction with independent Bernoulli randomization. This combination of randomization scheme and TATE estimator is totally ignorant of test-control interference. In order to do so, for both treatment A and treatment B we simulate 1,000 experiments in which, using Bernoulli randomization, 50\% of listings are assigned to treatment and 50\% of listings are assigned to control. For each simulated experiment, we use the difference-in-means estimator to estimate the TATE for both $BR$ and $ARPL$. The bottom rows of Figures \ref{fig:histogram_df_20} and \ref{fig:histogram_df_50} show the distribution of the estimated TATE for average revenue per listing and booking rate under the treatment A and treatment B, respectively. Compared to the ground truth, both pairs of TATE distributions are biased. As expected, the treatment effect estimates for booking rate overstate the effect of the treatment, as treated listings cannibalize bookings from control listings. In contrast, because a higher proportion of treated listings receive bookings, the treatment effect estimates for average revenue per listing understate the negative effect of the treatment. Table \ref{tab:ind_bias} reports the mean, standard deviation, bias, and RMSE for the naive difference-in-means estimator.

We compare these sampling distributions to those obtained using the naive difference-in-means estimator after determining treatment assignments using blocked graph cluster randomization. The top row of Figures \ref{fig:histogram_df_20} and \ref{fig:histogram_df_50} show the distribution of the TATE for average revenue per listing and booking rate under treatment A and treatment B, respectively. Table \ref{tab:gcr_bias} shows the mean, standard deviation, bias and RMSE for the four TATE esimator distributions generated under graph cluster randomization and block random assignment. Graph cluster randomization and block random assignment reduce the bias of the difference-in-means estimator relative to Bernoulli randomization. However, the variance of the TATE distributions \text{increases} relative to the distributions under Bernoulli randomization.

\begin{figure}[ht]
\begin{center}
\includegraphics[scale=.33]{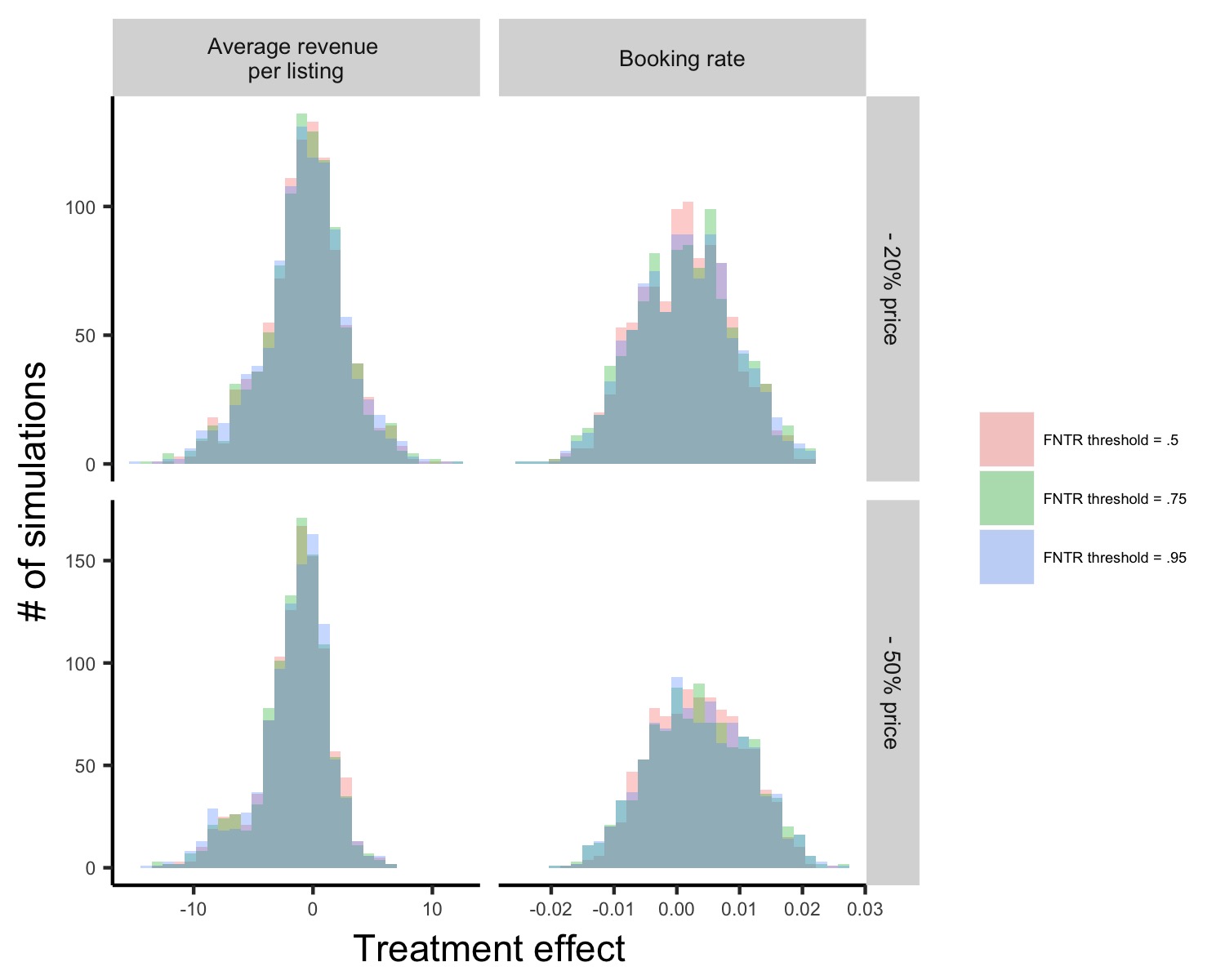}
\caption{The distribution of total average treatment effect estimates for booking rate and average revenue per listing across 1,000 simulations using graph cluster randomization and block random assignment, along with the FNTR exposure model. $\lambda$ values of .5, .75, and .95 are tested. The top row corresponds to treatment A, the bottom row corresponds to treatment B. The left column shows TATE estimates for average revenue per listing, the right column shows TATE estimates for booking rate.}
\label{fig:fntr_compare}
\end{center}
\end{figure}

We next consider pairing graph cluster randomization and block random assignment with the FNTR exposure model. Recall that the FNTR model has one hyperparameter, $\lambda$, which sets the threshold percentage of neighbors sharing the ego's treatment assignment at which the ego is included in the analysis. We test three different $\lambda$ values: .5, .75, and .95. Figure \ref{fig:fntr_compare} shows the TATE distributions corresponding to this design for each treatment intervention, outcome, and value of $\lambda$. Table \ref{tab:eff_bias} shows the mean, standard error, bias and RMSE for these TATE estimators. The impact of the FNTR exposure model is mixed. The effect of the FNTR model on bias is weakly positive for all values of $\lambda$; bias is either unchanged or decreases. The FNTR model also has a weakly positive impact on RMSE when $\lambda$ is small ($\lambda \in \{.5, .75\}$). However, for $\lambda = .95$, the FNTR model increases RMSE. This may indicate that high values of $\lambda$ result in a small sample that may not representative of the entire population. Even in cases where the FNTR model does decrease bias, the reduction is modest compared to the initial improvement from graph cluster randomization and block random assignment. Furthermore, the RMSEs achieved with the FNTR model are still large relative to the RMSE obtained with independent Bernoulli treatment assignment and the difference-in-means estimator.

\begin{figure}[ht]
\begin{center}
\includegraphics[scale=.33]{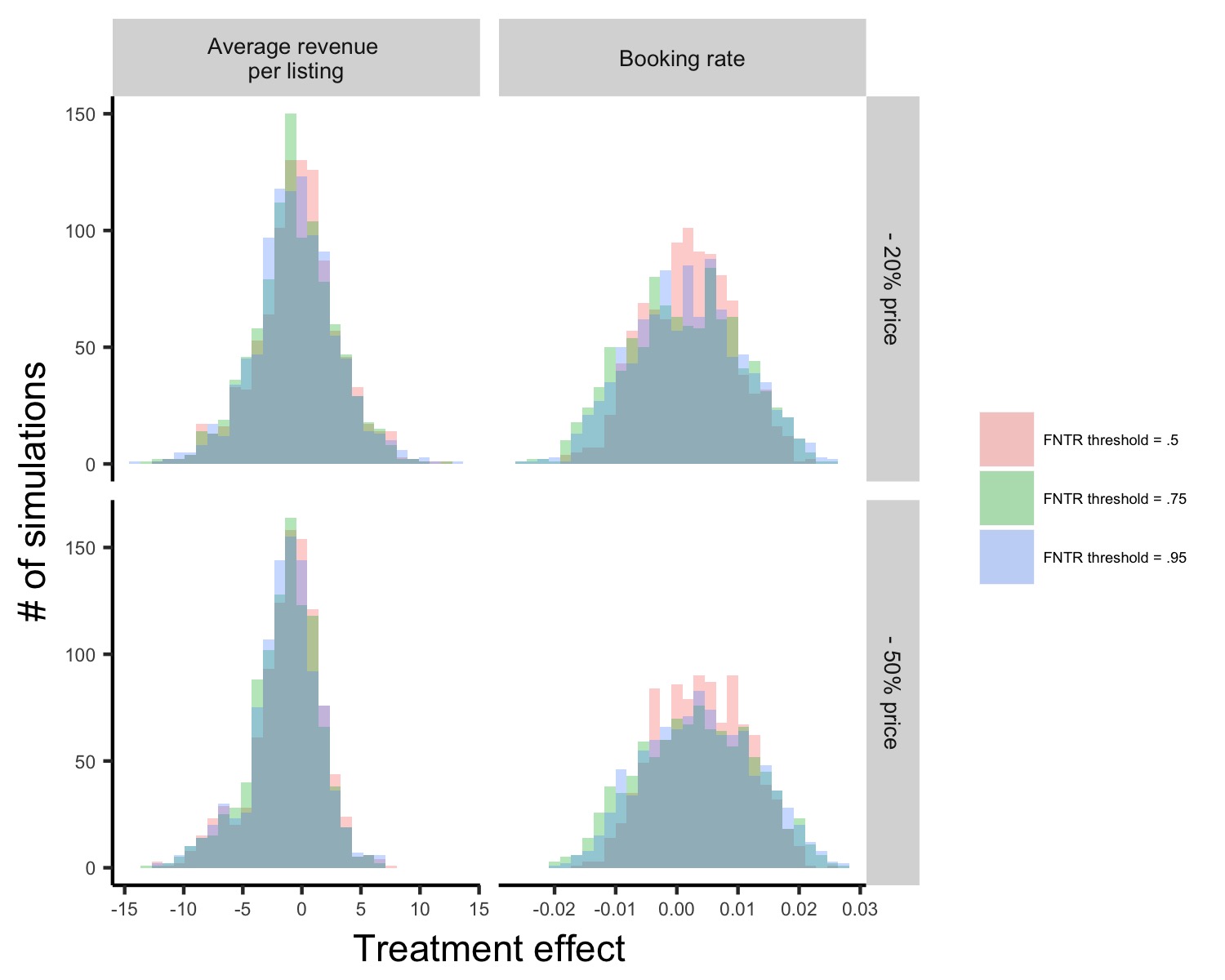}
\caption{The distribution of total average treatment effect estimates for booking rate and average revenue per listing across 1,000 simulations using graph cluster randomization and block random assignment, the FNTR exposure model, and the Hajek estimator. $\lambda$ values of .5, .75, and .95 are tested. The top row corresponds to treatment A, the bottom row corresponds to treatment B. The left column shows TATE estimates for average revenue per listing, the right column shows TATE estimates for booking rate.}
\label{fig:hajek_compare}
\end{center}
\end{figure}

Lastly, we consider a case that combines blocked graph cluster randomization, the FNTR exposure model, and the Hajek estimator. We test the same set of FNTR thresholds. Figure \ref{fig:hajek_compare} shows the TATE distributions corresponding to this design for each treatment intervention, outcome, and value of $\lambda$. Table \ref{tab:hajek_bias} shows the mean, standard error, bias and RMSE for these TATE estimators. The Hajek estimator has a neutral to negative effect on both bias and RMSE; in no case does the Hajek estimator produce bias or RMSE that is lower than those obtained using Bernoulli randomization and the difference in means estimator. This may be due to the finite sample bias of the Hajek estimator, in addition to potential issues with the size and representativeness of FNTR samples. Future work might compare the Hajek estimator to the Horvitz-Thompson estimator, as well as other, similar treatment effect estimators.

Overall, the combination of graph cluster randomization and block random assignment significantly reduces bias from test-control interference. However, this comes at the cost of increased estimator variance and RMSE. While the FNTR exposure model further reduced bias and RMSE, these improvements were modest compared to the initial gains from graph cluster randomization. Furthermore, the RMSE of the TATE distribution under the FNTR model remains higher than the RMSE of the TATE distribution that is maximally biased. We also find that the Hajek estimator has a negative effect on both bias and RMSE relative to the difference in means estimator.

\section{Discussion} \label{theory_extensions}

The fact that the methods we study decrease bias, but increase RMSE highlights an interesting tradeoff for marketplace designers and other practitioners, and suggests possible extensions to this work. While unbiased experiments are desirable, the corresponding loss of statistical power can lead to minimum detectable effect sizes that are much higher than acceptable. Large online marketplace firms often conduct A/B tests with the hopes of detecting treatment effects that are on the order of fractions of a percent, and these relatively small treatment effects can equate to gains or losses of millions of dollars. Given this fact, it may only make sense for firms to employ designs and estimators that reduce bias when there is a strong prior belief that the magnitude of bias from test-control interference will be large, and/or will flip the sign of a given experiment's treatment effect estimate. Future work might focus on identifying techniques that reduce bias without inducing such a large increase in treatment effect estimators' sampling variances, which would make methods that reduce bias in treatment effect estimates appealing to practitioners in a larger fraction of cases.

Although we argue that our simulation provides useful baseline for measuring the magnitude of bias in TATE estimates due to test-control interference, our simulation simplifies Airbnb's market dynamics in multiple ways. Future work might focus on making the simulation framework more sophisticated and realistic. For instance, the simulation framework could be modified to ensure that it is consistent with actual Airbnb data, or could be augmented to take into account general equilibrium effects. The simulation framework could also be extended to simulate more than one Airbnb calendar night, allow for Airbnb sellers to observe each others' price changes (and potentially mimic them), and to support intent-to-treat experimental interventions (i.e., treatment interventions that have seller non-compliance).

It may also be worthwhile to consider the robustness of the methods described in this paper to network misspecification, and to study the extent to which these methods generalize to other, similar settings. In this study, we are able to cluster Airbnb listings based on the same listing-level covariates used by both search algorithms and searchers. However, researchers may not know \textit{ex ante} which listing-level attributes best capture marketplace competition. When the ``wrong" set of covariates are used by researchers to generate a network of listings, the methods we present may not be as effective at reducing the bias of TATE estimators. One relatively straightforward approach to determine to what extent these methods are robust to network misspecification would be to measure their efficacy after injecting ``noise" into the network, e.g., through random edge creation, deletion, or rewiring. It would also be worthwhile to use simulations to measure the effectiveness of the methods described in this paper across other Airbnb markets, and in other marketplaces where, in addition to substitution, complementarity between items is more common. In the presence of both substitution and complementarity, it is not immediately clear how test-control interference may bias different TATE estimates.

Finally, it would be exciting to test the experiment design and analysis techniques discussed in this paper in the field. One approach to doing so would be to conduct an online meta-experiment, similar to those described by \citet{pouget2017testing} and \citet{saveski2017detecting}. In such a meta-experiment, two field experiments are conducted simultaneously; while each experiment has the same treatment intervention, the two experiments have different designs. By comparing the treatment effect estimates obtained in each ``meta-treatment arm" of the meta-experiment, it is possible to compare the treatment effect estimate obtained under the two experiment designs in a statistically rigorous way.

\section{Conclusion} \label{conclusions}

Given the ubiquity of large, valuable online marketplace firms, such as Alibaba, Amazon, Airbnb, and Uber, it is crucial to develop methods for obtaining accurate causal estimates through experimentation in online marketplaces. In this work, we have used a simulation framework to test the efficacy of various experiment design and analysis techniques in reducing bias in online marketplace experiments due to test control interference. Our simulations show that the combination of graph cluster randomization and block random assignment significantly reduces bias from test-control interference. However, this comes at the cost of increased RMSE. While the FNTR exposure model further reduced bias, these additional gains were modest compared to the initial bias reduction coming from GCR. Importantly, even our least biased TATE estimator still has a much higher RMSE than the original, maximally biased TATE estimator. These results indicate that methods from the network experimentation literature can be effective at reducing bias from test-control interference in online marketplace settings, however, the bias reductions we observe come at the cost of much higher treatment effect estimator sampling variance. This trade-off may not be worthwhile for firms that conduct A/B tests with the hopes of detecting treatment effects that are, in relative terms, small, but can correspond to gains or losses of millions of dollars. In light of this trade-off, we recommend a future research agenda that focuses on increasing the sophistication and realism of the simulation framework described in this paper, extending this paper's simulation framework to encompass a wider variety of online marketplace settings, and/or conducting in-vivo meta-experiments to measure the efficacy of the methods we propose in the field.

\clearpage
\section{Figures}

\begin{figure}[ht]
	\centering
	\includegraphics[scale=.33]{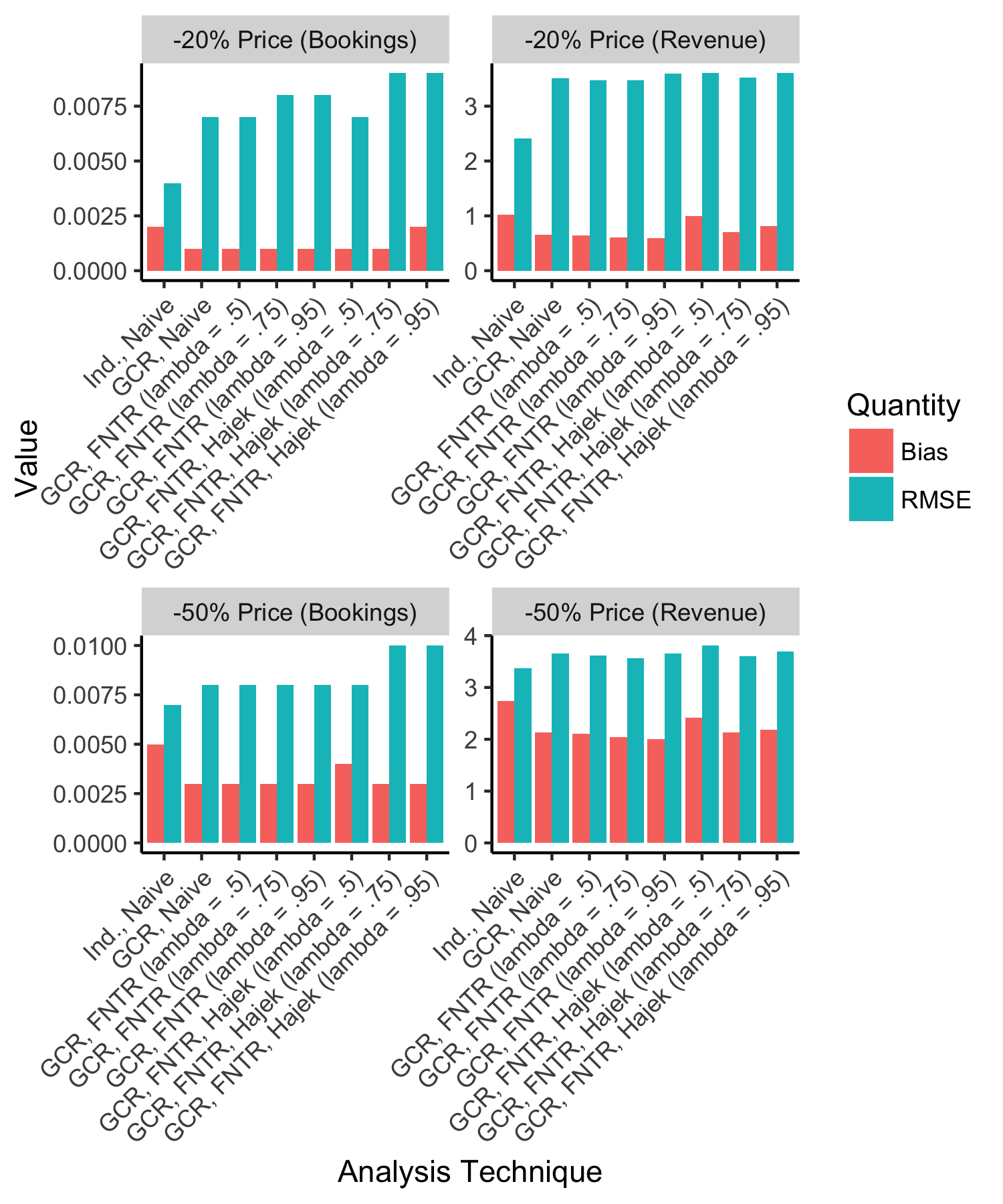}
	\caption{Changes in bias and RMSE of TATE estimates using different experiment designs and TATE estimators. Top panels correspond to treatment A. Bottom panels correspond to treatment B. Left panels correspond to booking rate TATE. Right panels correspond to average revenue per listing TATE. 
}
	\label{fig:overall_comparison}
\end{figure}

\clearpage
\section{Tables}

\begin{table}[ht]
    \begin{center}
        \caption{Distribution of Airbnb Listing Room Types}\label{tab:listing_room_types}
        \begin{tabular}{cccc}\hline
        & Entire home/apt & Private room & Shared room \\
        \hline
        Count & 6,566 & 2,062 & 227 \\ \hline
        \end{tabular}
    \end{center}
\end{table}

\begin{table}[ht]
   \begin{center}
   \caption{Distribution of Airbnb Listing Covariates} \label{tab:listing_cov}
   \begin{tabular}{c|lllllll} \hline
   & Reviews & Satisfaction & Capacity & Beds & Baths & Price & Min Stay \\ \hline
   Min & 0 & 1.00 & 1 & 0 & 0 & \$15 & 1 \\
   1st Qu. & 0 & 4.50 & 2 & 1 & 1 & \$89 & 1 \\
   Median & 3 & 4.50 & 3 & 1 & 1 & \$140 & 2 \\
   Mean & 11.40 & 4.59 & 3.06 & 1.40 & 1.37 & \$226 & 3.29 \\
   3rd Qu. & 12 & 5.00 & 4 & 2 & 2 & \$249 & 3 \\
   Max & 304 & 5.00 & 8 & 10 & 8 & \$10,000 & 365 \\
   N/A & 0 & 2,422 & 2,226 & 12 & 933 & 0 & 437 \\ \hline
   \end{tabular}
  \end{center}
\end{table}

\begin{table}[ht]
   \begin{center}
   \caption{True counterfactual outcome distributions and TATEs} \label{tab:compare_baseline}
   \begin{tabular}{llllllll} \hline
   & $\mu_C$ & $\sigma_C$ & $\mu_T$ & $\sigma_T$ & $t$ & $p$ & TATE \\ \hline
   -50\% (revenue) & \$7.25 & \$2.26 & \$3.65 & \$1.13 & 45.86 & $< 2.2 \times 10^{-16}$ & -\$3.60 \\
   -50\% (bookings) & 0.024 & 0.002 & 0.024 & 0.002 & 1.12 & 0.20 & 0 \\
   -20\% (revenue) & \$7.25 & \$2.26 & \$5.93 & \$1.86 & 14.54 & $< 2.2 \times 10^{-16}$ & -\$1.33 \\
   -20\% (bookings) & 0.024 & 0.002 & 0.024 & 0.002 & 2.17 & 0.03 & $1.05 \times 10^{-5}$ \\ \hline
   \end{tabular}
  \end{center}
\end{table}

\begin{table}[ht]
   \begin{center}
   \caption{TATE mean, standard deviation, bias, and RMSE under random assignment and difference-in-means} \label{tab:ind_bias}
   \begin{tabular}{lllll} \hline
   & $\mu_{ATE}$ & $\sigma_{ATE}$ & Bias & RMSE \\ \hline
   -50\% (revenue) & -\$0.86 & \$1.95 & \$2.74 & \$3.37\\
   -50\% (bookings) & 0.005 & .003 & 0.005 & 0.007 \\
   -20\% (revenue) & -\$0.31 & \$2.18 & \$1.02 & \$2.41\\
   -20\% (bookings) & 0.002 & .003 & 0.002 & 0.004 \\ \hline
   \end{tabular}
  \end{center}
\end{table} 

\begin{table}[ht]
   \begin{center}
   \caption{TATE mean, standard deviation, bias, and RMSE under graph cluster assignment, black random assignment, and difference-in-means} \label{tab:gcr_bias}
   \begin{tabular}{lllll} \hline
   & $\mu_{ATE}$ & $\sigma_{ATE}$ & Bias & RMSE \\ \hline
   -50\% (revenue) & -\$1.47 & \$2.97 & \$2.13 & \$3.65\\
   -50\% (bookings) & 0.003 & 0.007 & 0.003 & 0.008 \\
   -20\% (revenue) & -\$0.68 & \$3.45 & \$0.65 & \$3.51 \\
   -20\% (bookings) & 0.001 & 0.007 & 0.001 & 0.007 \\ \hline
   \end{tabular}
  \end{center}
\end{table} 

\begin{table}[ht]
   \begin{center}
   \caption{TATE mean, standard deviation, bias, and RMSE under graph cluster assignment, block random assignment, and difference-in-means for units in the effective treatment or effective control (FNTR)} \label{tab:eff_bias}
   \begin{tabular}{llllll} \hline
   & $\lambda$ & $\mu_{ATE}$ &  $\sigma_{ATE}$ & Bias & RMSE \\ \hline
   -50\% (revenue) & 0.5 & -\$1.49 & \$2.92 & \$2.11 & \$3.61\\
   -50\% (bookings) & 0.5  & 0.003 & 0.007 & 0.003 & 0.008 \\
   -20\% (revenue) & 0.5 & -\$0.69 & \$3.41 & \$0.64 & \$3.47 \\
   -20\% (bookings) & 0.5  & 0.001 & 0.007 & 0.001 & 0.007 \\ 
   -50\% (revenue) & 0.75  & -\$1.56 & \$2.93 & \$2.04 & \$3.57\\
   -50\% (bookings)  & 0.75 & 0.003 & 0.008 & 0.003 & 0.008 \\
   -20\% (revenue)  & 0.75  & -\$0.72 & \$3.41 & \$0.61 & \$3.47 \\
   -20\% (bookings)  & 0.75  & 0.001 & 0.008 & 0.001 & 0.008 \\ 
   -50\% (revenue)  & 0.95  & -\$1.60 & \$3.05 & \$2.00 & \$3.65\\
   -50\% (bookings)  & 0.95  & 0.003 & 0.008 & 0.003 & 0.008 \\
   -20\% (revenue)  & 0.95  & -\$0.74 & \$3.54 & \$0.59 & \$3.59 \\
   -20\% (bookings) & 0.95  & 0.001 & 0.008 & 0.001 & 0.008 \\ \hline
   \end{tabular}
  \end{center}
\end{table} 

\begin{table}[ht]
   \begin{center}
   \caption{TATE mean, standard deviation, bias, and RMSE under graph cluster assignment and block random assignment using the Hajek estimator for units in the effective treatment or effective control (FNTR)} \label{tab:hajek_bias}
   \begin{tabular}{llllll} \hline
   & $\lambda$ & $\mu_{ATE}$ &  $\sigma_{ATE}$ & Bias & RMSE \\ \hline
   -50\% (revenue) & 0.5 & -\$1.19 & \$2.94 & \$2.41 & \$3.81\\
   -50\% (bookings) & 0.5  & 0.004 & $5.4 \times 10^{-5}$ & 0.004 & 0.008 \\
   -20\% (revenue) & 0.5 & -\$0.933 & \$3.45 & \$1.00 & \$3.60 \\
   -20\% (bookings) & 0.5  & 0.001 & $5.4 \times 10^{-5}$ & 0.001 & 0.007 \\ 
   -50\% (revenue) & 0.75  & -\$1.47 & \$2.90 & \$2.13 & \$3.60\\
   -50\% (bookings)  & 0.75 & 0.003 & 0.009 & 0.003 & 0.01 \\
   -20\% (revenue)  & 0.75  & -\$0.63 & \$3.45 & \$0.70 & \$3.52 \\
   -20\% (bookings)  & 0.75  & 0.001 & $8.7 \times 10^{-5}$ & 0.001 & 0.009 \\ 
   -50\% (revenue)  & 0.95  & -\$1.41 & \$2.97 & \$2.19 & \$3.69\\
   -50\% (bookings)  & 0.95  & 0.003 & 0.009 & 0.003 & 0.01 \\
   -20\% (revenue)  & 0.95  & -\$0.52 & \$3.51 & \$0.81 & \$3.60 \\
   -20\% (bookings) & 0.95  & 0.002 & 0.009 & 0.002 & 0.009 \\ \hline
   \end{tabular}
  \end{center}
\end{table} 

\clearpage

\ACKNOWLEDGMENT{%
We thank Dean Eckles, Ed McFowland, Rajiv Mukherjee, and Michael Zhao for their useful feedback. We also thank participants at the 2016 Conference on Digital Experimentation, the 2016 Workshop on Information Systems and Economics, and the 2019 Workshop on Experimental and Behavioral Economics in Information Systems for their comments.
}

%
%
%


\bibliographystyle{informs2014} 
\bibliography{main.bib} 


\end{document}